\documentclass[twocolumn,aps,prd,showpacs, nofootinbib]{revtex4-2}
\pdfoutput=1
\usepackage{amsmath,amsfonts}
\usepackage{epsfig}
\usepackage{caption}
\usepackage{subcaption}
\usepackage{ifpdf}
\usepackage{hyperref}
\usepackage{amsmath}
\usepackage{graphicx}
\usepackage{color}
\usepackage{natbib}
\usepackage{multirow}
\usepackage{kotex}
\usepackage{times}

\newcommand{\msun}{M_\odot}
\newcommand{\be}{\begin{equation}}
\newcommand{\ee}{\end{equation}}
\newcommand{\bea}{\begin{eqnarray}}
\newcommand{\eea}{\end{eqnarray}}

\newcommand{\etal}{{\it et al.}}

\begin{document}

\title{Systematic bias due to eccentricity in parameter estimation for merging binary neutron stars : Spinning case}

\author{Eunjung Lee}
\affiliation{Department of Physics, Pusan National University, Busan, 46241, Korea}

\author{Hee-Suk Cho}
\email{chohs1439@pusan.ac.kr}
\affiliation{Department of Physics, Pusan National University, Busan, 46241, Korea}
\affiliation{Extreme Physics Institute, Pusan National University Busan, 46241, Korea}

\author{Chang-Hwan Lee}
\affiliation{Department of Physics, Pusan National University, Busan, 46241, Korea}

\date{\today}

\begin{abstract}
In our previous work [Phys. Rev. D {\bf 105}. 124022 (2022)],
we studied the impact of eccentricity on gravitational-wave parameter estimation for a nonspinning binary neutron star (BNS) system.
We here extend the work to a general binary system by including the spin parameter.
As in the previous work, we employ the analytic Fisher-Cutler-Vallisneri method to calculate
the systematic bias that can be produced by using noneccentric waveforms in parameter estimation,
and we verify the reliability of the method by comparing it with numerical Bayesian parameter estimation results.
We generate $10^4$ BNS sources randomly distributed in the parameter space $m_1$--$m_2$--$\chi_{\rm eff}$--$e_0$,
where the neutron star mass is in the range of $1\msun \leq m_{1,2}\leq  2\msun (m_2 \leq m_1)$, the effective spin is $-0.2 \leq \chi_{\rm eff} \leq0 .2$,
and the eccentricity (at the reference frequency 10 Hz) is $0 \leq e_0 \leq 0.024$.
For the true value of the tidal deformability ($\lambda$) of neutron stars, we assume the equation of state model APR4.
For all gravitational-wave signals emitted from the sources, we calculate the systematic biases ($\Delta \theta$)
for the chirp mass ($M_c$), symmetric mass ratio ($\eta$), effective spin ($\chi_{\rm eff}$), and effective tidal deformability ($\tilde{\lambda}$),
and obtain generalized distributions of the biases.
The distribution of biases in $M_c, \eta$, and $\chi_{\rm eff}$ shows narrow bands, 
and the median value of the bias increases or decreases quadratically with increasing $e_0$,
indicating a weak dependence of biases on the three parameters. 
On the other hand, the biases of $\tilde{\lambda}$ are widely distributed depending on the values of the mass and spin parameters at a given $e_0$.
We investigate the implications of biased parameters for the inference of neutron star properties by performing Bayesian parameter estimation for specific cases.
We find that a BNS signal consisting of two neutron stars within the typical mass range $[1,2]\msun$ can be estimated to be
a BNS signal whose component mass is much smaller or much larger than the typical mass range.
In particular, by showing concrete examples, we  demonstrate that parameter estimation using noneccentric waveforms for eccentric
BNS signals can yield false predictions for the neutron star equation of state.

\end{abstract}

%\pacs{04.30.--w, 04.80.Nn, 95.55.Ym}

\maketitle
%=======	Title		================================	

%=======	Intro		================================	
\section{Introduction}
Recently, the LIGO-Virgo-KAGRA (LVK) announced the discovery of the 200th gravitational wave (GW) signal \cite{LIGOVirgoKAGRAAnnounce200th}. 
This signal is believed to be emitted from a binary black hole (BBH) merger. 
The number of GW observations has increased dramatically as the GW detectors have achieved technological improvements. 
While most GW signals were produced by BBHs, several candidates were predicted as binary neutron star (BNS)  or neutron star-black hole (NSBH) systems. 
In 2023, a GW signal consisting of a neutron star (NS) and a compact object whose mass is within the mass gap was also observed, 
which has gained considerable attention in astronomy \cite{LIGOVirgoFindsMystery, Abac_2024}.
The 3G GW detectors, Einstein Telescope \cite{Branchesi_2023} and Cosmic Explorer \cite{galaxies10040090}, are aiming to operate in the mid 2030s,
and their network will significantly increase the detection rate and allows for much longer observation time of GW signals.

NS properties such as mass, radius, and tidal deformability are mainly governed by the equation of state (EoS), 
and theoretical nuclear physicists have established various EoS models over the past few decades.
Electromagnetic observations of NSs, including radio and x-ray data, have also been made 
to constrain NS properties. In particular, simultaneous measurements of mass and radius for pulsars using the 
Neutron Star Interior Composition Explorer (NICER)  
instrument have recently made significant progress in constraining NS EoS \cite{Miller_2019,Jiang_2020,Pang_2021,Raaijmakers_2021,Rutherford_2024,MALIK2025100086}.
Finally, the successful detection of GW170817 and its electromagnetic counterparts has opened up multimessenger astronomy beyond the optical limit \cite{Abbott_2017,Abbott_2017b,Abbott_2017c,Radice_2018,BAUSWEIN2019167958}.
GW170817, the first detected BNS GW signal,  provided a constraint on the value of tidal deformability,
and the results showed that NS EoS favors soft EoS models \cite{PhysRevLett.119.161101, PhysRevX.9.011001,PhysRevLett.121.161101}.

Eccentricity is one of the intrinsic physical parameters of merging binaries, but this parameter has been ignored in past GW searches and parameter estimation
because merging binaries are likely to settle into quasicircular orbits when they enter the detector sensitivity frequency band.
However, it has been shown that dynamically formed BBHs in dense stellar clusters can still have eccentricities larger than 0.1 
at 10 Hz \cite{PhysRevD.97.103014, PhysRevD.97.103014}.
Recently, some efforts have been made to find eccentric signals in the LVK data \cite{Nitz_2020,Nitz_2021,PhysRevLett.127.151101,PhysRevD.111.103018}, 
and to determine the eccentricity contained in the GW signals observed so far \cite{10.1093/mnras/staa2120,Morras:2025xfu,cv75-y8dr,Gupte:2024jfe}.
Several works showed how the detectability decreases due to ignoring eccentricity in GW searches \cite{PhysRevD.110.044013,PhysRevD.111.043040,PhysRevD.109.043037}.
It has also been shown that ignoring eccentricity can yield systematic biases when estimating source parameters \cite{PhysRevD.109.043037, mc64-1jtd}.  
Recently, Roy and Saini \cite{DuttaRoy:2024aew} studied the systematic errors due to unmodeled eccentricity in the tidal deformability measurement 
and showed that the systematic errors exceed the statistical errors at an eccentricity of $\sim 10^{-3}$ at 10 Hz reference GW frequency for
the 3 G detectors, Cosmic Explorer and Einstein Telescope.  
They briefly presented a trend between statistical and systematic errors according to eccentricity, considering a single fiducial equal-mass BNS source with $m_i=1.4\msun$,
In this work, we also study the systematic errors due to eccentricity employing the same method as in \cite{DuttaRoy:2024aew}.
However,  we consider various NS masses and spins and assume the aLIGO detector sensitivity.
We will show that our result is consistent with that of \cite{DuttaRoy:2024aew} when considering the same BNS source and detector sensitivity.

In our previous work \cite{Cho:2022cdy}, we studied the impact of eccentricity on parameter estimation for nonspinning BNSs.
(Several works related to this issue are summarized in \cite{Cho:2022cdy}.)
Here, we extend the work to a spinning system to reflect more realistic GW signals and employ the same methodology as in \cite{Cho:2022cdy}.
In particular, we focus on the tidal parameter and investigate how systematic biases can affect the NS EoS estimation.

\section{Method}
In this section, we briefly summarize our methodology (for more details, refer to the previous work \cite{Cho:2022cdy}).

\subsection{TaylorF2}
 We use the TaylorF2 waveform model given by \cite{PhysRevD.80.084043}
 \begin{align}
h(f) = A f^{-7/6} \, e^{i \Psi(f)},
\label{eq:TalyorF2}
\end{align}
where $A$ is the GW amplitude.
The GW phase is given by
\bea   \label{eq:TaylorF2phase}
\Psi(f)&=&2\pi ft_c -2 \phi_c -{\pi\over 4}+\frac{3}{128 \eta v^5} [\Psi^{\rm pp,circ}(f) \\     \nonumber
 &+& \Psi^{\rm spin}(f)+ \Psi^{\rm tidal}(f)+\Psi^{\rm ecc}(f)],
\eea
where $t_c$ and $\phi_c$ are the coalescence time and phase, respectively 
and $v \equiv [\pi  (m_1 + m_2)f]^{1/3}$ = \( (\pi M f)^{1/3} \) is the post-Newtonian (PN) expansion parameter 
that characterizes the orbital velocity of the binary. 
In the bracket, $\Psi^{\rm pp, circ}$ is  the 3.5PN point-particle circular correction term, 
$\Psi^{\rm spin}$ is  the 3.5PN spin correction term, $\Psi^{\rm tidal}$ is the 6PN  tidal correction term, 
and $\Psi^{\rm ecc}$ is the 3PN eccentricity correction term, respectively. Note that, except $\Psi^{\rm spin}$, 
all terms are the same as in the previous work \cite{Cho:2022cdy}.
Full expressions of $\Psi^{\rm pp, circ}$ and $\Psi^{\rm spin}$ are given in Ref. \cite{PhysRevD.80.084043}.
The tidal term is given by \cite{PhysRevLett.112.101101,PhysRevD.91.043002}
\be \label{eq.tidal correction}
 \Psi^{\rm tidal} = - \bigg[ \frac{39 \tilde{\lambda}}{2}v^{10} + \bigg( \frac{3115\tilde{\lambda}}{64} - \frac{6595 \sqrt{1-4\eta} \delta \tilde{\lambda}}{364} \bigg)v^{12} \bigg],
\ee
where $\tilde{\lambda}$ and  $\delta \tilde{\lambda}$ are effective tidal parameters determined by the component tidal deformabilities $(\lambda_1, \lambda_2)$.
The eccentricity term is given in Eq. (6.26) of Ref. \cite{PhysRevD.93.124061}
\bea
\Psi^{\rm ecc}&=&-\frac{2355}{1462} e_0^2 \left(\frac{v_0}{v}\right)^{19/3} \bigg[1+\bigg(\frac{18766963}{2927736}\eta \\   \nonumber
&+&\frac{299076223}{81976608}\bigg) v^2 +\left(\frac{2833}{1008}-\frac{197}{36} \eta \right) v_0^2  \\   \nonumber
&+& O(v^4)+...+O(v^6)\bigg],
\eea
where $e_0$  is the value of eccentricity at the reference frequency $f_0$, and the choice of $f_0$ is arbitrary.
While the effect of eccentricity appears at very low orders in the PN expansion, 
in practice, it is often ignored because most binary events are almost circular at the time of observation. 
However, in the low frequency region, the effect of eccentricity on the bias becomes large 
because the eccentricity may still remain ($e_0 \gtrsim 0.01$) \cite{PhysRevD.93.124061}.

\subsection{Systematic bias in parameter estimation}
The Bayesian parameter estimation and the Fisher matrix (FM) methods are based on matching the model waveforms to observational data.
In the Bayesian method, matching is performed over the entire parameter space, 
while in the FM method, it is performed at the true parameter position 
and the result is obtained through Gaussian approximation error prediction at high SNR \cite{Thrane:2018qnx}.
The match is obtained from the inner product defined by
\begin{align}
\langle x \mid h \rangle = 4 \, \text{Re} \int_{f_{\rm min}}^{f_{\rm max}} \frac{\tilde{x}(f) \, \tilde{h}^*(f)}{S_n(f)} \, df,
\label{eq:inner_product}
\end{align}
where $x$ is observed data, $h$ is a model waveform, and  $S_n$ is the noise power spectral density (PSD) of the detector.
Through the Bayesian analysis, the parameter estimation result is given by the posterior distribution functions (PDFs) for the parameters $\theta_i$
\begin{align}
p(\theta \mid x) \propto p(\theta) \cdot L(x \mid \theta),
\end{align}
where $p(\theta)$ refers to the prior distribution for a given parameter $\theta$, which is usually given as a flat distribution within physical boundaries. 
Therefore, PDFs are mainly determined by the likelihood $L$, and assuming high SNRs, the final form of $L$ is given by \cite{PhysRevD.49.2658, PhysRevD.46.5236, Cho13}
\begin{equation}
L(\theta) \propto \exp\left[ -\rho^2 \left\{ 1 - \langle \hat{s} \mid \hat{h}(\theta) \rangle \right\} \right],
\label{eq:normalized}
\end{equation}
where $s$ is the signal, $\rho = \sqrt{ \langle s \mid s \rangle }$ is the SNR, and  $\hat{h}$ means normalization of $h$, i.e., $\langle \hat{h} \mid \hat{h} \rangle = 1$. 
Consequently,  the shape of $L$ is determined through the difference between the observed signal and the model waveforms, and the SNR mainly determines the sharpness of $L$.
If the waveform model $h$ is perfectly correct, the likelihood has a peak when the parameter values of $h$ are the same as the signal's parameter values.
In principle, due to the incompleteness of the model, the recovered parameter values cannot be the same as the true values,
and the difference (i.e., systematic bias) depends on the accuracy of the model.

If the SNR is high enough, the measurement error can be calculated approximately by using the FM approach.
The FM is given by
\begin{equation}
\Gamma_{ij} = \left\langle \frac{\partial h}{\partial \theta_i} \middle| \frac{\partial h}{\partial \theta_j} \right\rangle \bigg|_{\theta = \theta_{\text{true}}},
\end{equation}
where, $\theta_{\text{true}}$ is the true value.
The covariance matrix can be given by an inversion matrix of FM ($\Sigma = \Gamma^{-1}$),
and the measurement error $\sigma_i$ and the correlation coefficient $C_{ij}$ can be obtained as
\begin{equation}
\sigma_i = \sqrt{\Sigma_{ii}}, \quad C_{ij} = \frac{\Sigma_{ij}}{\sqrt{\Sigma_{ii} \Sigma_{jj}}}.
\label{eq:sigma}
\end{equation}

While the Bayesian method is free to apply the priors, only Gaussian priors can be applied analytically to the FM method.
The prior-incorporated covariance can be obtained just by adding the component $\Gamma^0_{ii}\equiv (P_{\theta_i})^{-2}$ in the following way \cite{PhysRevD.49.2658, PhysRevD.52.848}, 
\begin{equation}
\Sigma_{ij}^P = \left( \Gamma_{ij} +  \Gamma^0_{ii} \right)^{-1}.
\label{eq:Sigma}
\end{equation}
where $P_{\theta_i}$ indicates the standard deviation of the Gaussian prior for the parameter $\theta_i$.
Note that Eqs. \eqref{eq:sigma} and \eqref{eq:Sigma} do not follow standard Einstein notation for indices.
Based on the FM formalism, Cutler and Vallisneri \cite{PhysRevD.76.104018} developed an analytic method (henceforth
denoted as“FCV”) to calculate the systematic bias. 
In this work, the bias can be given by \cite{PhysRevLett.112.101101}
\begin{equation}
\Delta \theta_i = 4A^2 \Sigma^P_{ij} \int_{f_{\min}}^{f_{\max}} \frac{f^{-7/3}}{S_n} \left( \Psi_T - \Psi_{\text{AP}} \right) \partial_j \Psi_{AP} \, df,
\end{equation}
where $A$ indicates the amplitude in Eq. \eqref{eq:TalyorF2}, $\Psi_{\rm T}$ is the true GW phase given in Eq. \eqref{eq:TaylorF2phase},
and $\Psi_{\rm AP}$ is the approximate phase that omits the eccentricity term.

\section{Result}

\subsection{Setup}
We use the eccentric waveform model ``TaylorF2Ecc" imlemented in LIGO Algorithm Library (LAL) \cite{lalsuite}
and adopt the aLIGO PSD labeled as $\mathrm{aligo\_O4high}$ \cite{LIGOT2000012v2NoiseCurves}.
The frequency range is set from $f_{\rm min}=20$ to $f_{\rm max}=f_{\rm ISCO}=1/[6^{3/2}\pi M]$  , where ISCO means the innermost stable circular orbit.  
As our reference BNS source, we choose a binary with the true values of $(m_1,m_2, [M_c,\eta],\chi_{\rm eff},\tilde{\lambda},\delta \tilde{\lambda })=(2\msun,1\msun,[1.21673 \msun,0.22222],0.05,237,101)$, where we apply the APR4 EoS model \cite{PhysRevD.81.123016} for the tidal parameters 
\footnote{Although a ($2\msun, 1\msun$) BNS is astrophysically quite
unlikely, we choose this binary for consistency with our previous work \cite{Cho:2022cdy}.}.

As in the previous work \cite{Cho:2022cdy}, we perform Bayesian parameter estimation using 
BILBY \cite{Ashton:2018jfp}, DYNESTY \cite{10.1093/mnras/staa278}, and a multibanded likelihood technique \cite{PhysRevD.104.044062} for BNS injection signals. 
We assume flat priors in the ranges of $[0.5 \msun, 3\msun]$ for the masses $(m_1, m_2)$, 
$[0,0.99]$ for the spins $(\chi_1, \chi_2)$, and $[0, 5000]$ for the component tidal parameters $(\lambda_1, \lambda_2)$.
In general, the priors of $t_c$ and $\phi_c$ are given as $[t_{c0}-1s,t_{c0}+1s]$ and $[0, 2\pi]$, respectively.
The extrinsic parameters are fixed because they have minimal effect on the intrinsic parameters (refer to \cite{Cho_2014,Cho_2022,Cho:2022cdy} for more details with concrete examples).
Consequently, the algorithm explores 8-d parameter space; $(M_c,\eta,\chi_1,\chi_2,\lambda_1, \lambda_2,\phi_c,t_c)$.
When displaying the PDFs in BILBY, we use the chirp mass $M_c \equiv (m_1 m_2)^{3/5}/M^{1/5}$ and the symmetric mass ratio $\eta \equiv m_1 m_2/M^2$ rather than $(m_1, m_2)$,
the effective spin $\chi_{\rm eff} \equiv (m_1 \chi_1+ m_2 \chi_2)/M$ rather than $(\chi_1, \chi_2)$,
and $(\tilde{\lambda},\delta \tilde{\lambda})$ rather than $(\lambda_1, \lambda_2)$.

In FM and FCV calculations, only the GW phase in Eq. \eqref{eq:TaylorF2phase} is important because the amplitude only scales the SNR. 
In this work, the FM consists of the seven parameters $(M_c,\eta,\chi_{\rm eff},\tilde{\lambda},\delta \tilde{\lambda},\phi_c,t_c)$.
We incorporate Gaussian priors; $P_{\phi_{c}} = \pi$, $P_{\chi_{\rm eff}} = 1$, and $P_{\delta\tilde{\Lambda}} = 100$ (the choice of $P_{\delta\tilde{\Lambda}}$ will be discussed later), respectively, and assume no priors for $M_c,\eta, \tilde{\lambda},$ and $t_c$.

\subsection{Validation of FCV method: Comparison with Bayesian parameter estimation}

\begin{figure*}[t]
\begin{center}
\includegraphics[width=2 \columnwidth]{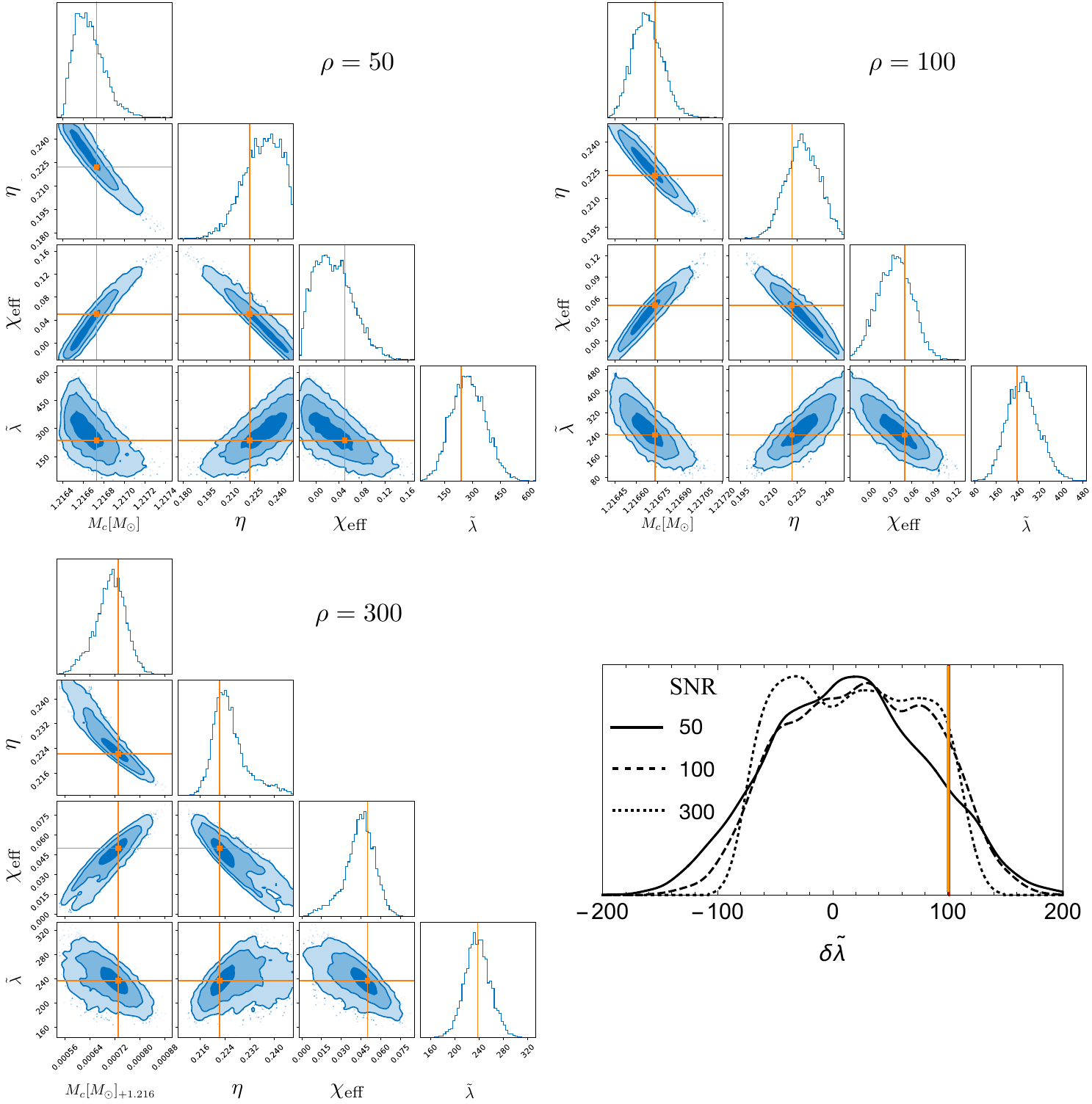}
\caption{\label{fig.bilby-pdfs}  Bayesian parameter estimation results calculated with various SNRs. The contours indicate 39, 86, and 99$\%$ confidence regions.
Injection values are  $(m_1,m_2, [M_c,\eta],\chi_{\rm eff},\tilde{\lambda},\delta \tilde{\lambda })=(2\msun,1\msun,[1.21673 \msun,0.22222],0.05,237,101)$ marked in orange.
PDFs for $\delta\tilde{\Lambda}$ are shown in the bottom right, separately.}
\end{center}
\end{figure*}

\begin{figure}[t]
\begin{center}
\includegraphics[width=\columnwidth]{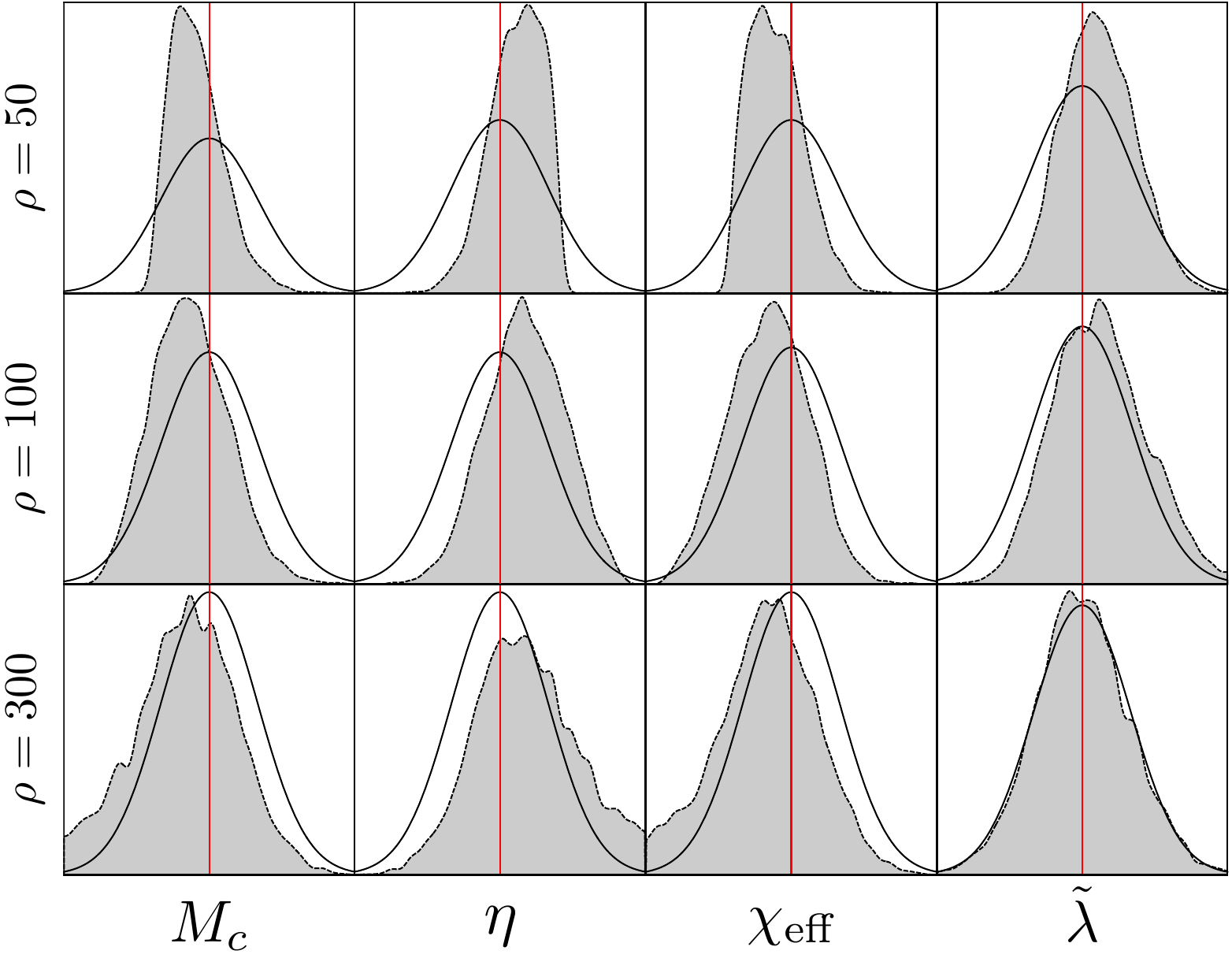}
\caption{\label{fig.bilby-Fisher-comparison} Comparison between the FM (black) and
Bayesian parameter estimation (gray). All parameters are successfully recovered overall, but those are biased at low SNRs. }
\end{center}
\end{figure}

Since the FCV approach is based on the FM method, it is important to calculate the FM accurately for the BNS system, and the validity of the FM can be examined by comparing the FM results with the Bayesian parameter estimation results.
For this purpose, we perform Bayesian parameter estimation for our fiducial BNS signal assuming SNRs of 50, 100, and 300, respectively.
Here, we use noneccentric waveforms for both the signal and the templates.
Figure \ref{fig.bilby-pdfs} shows the results for the parameters $(M_c, \eta, \chi_{\rm eff}, \tilde{\lambda},\delta \tilde{\lambda})$,
where the true values are marked in orange.
In all cases, the mass and spin parameters are biased at low SNRs, but all parameters are successfully recovered overall.
On the other hand, the PDFs for $\delta \tilde{\lambda}$ show similar shapes centered around 0 for all SNRs,
and this behavior is consistent with the nonspinning case in our previous work \cite{Cho:2022cdy}.

\begin{figure}[t]
\includegraphics[width=0.8\columnwidth]{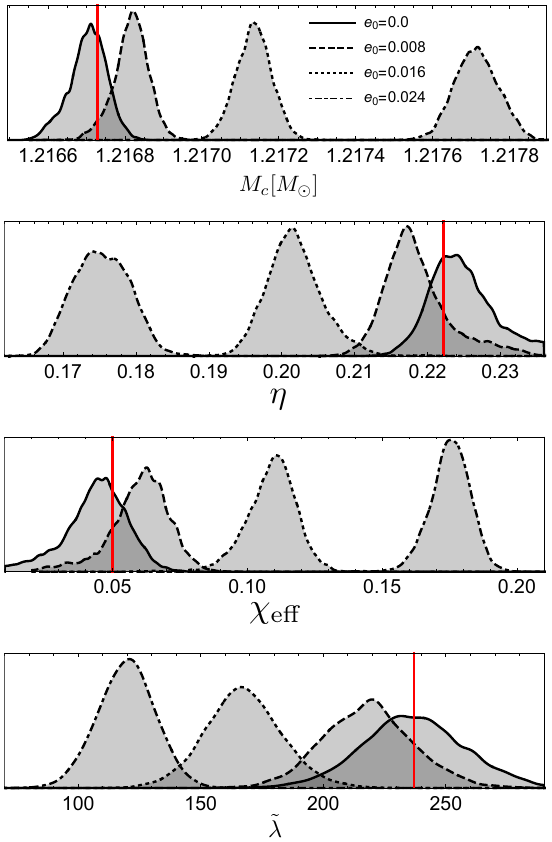}
\caption{\label{fig.bilby-1d-bias} Bayesian PDFs biased due to eccentricity. The true value is marked in red. We assume $\rho=300$.}
\end{figure}

\begin{figure}[t]
\begin{center}
\includegraphics[width=\columnwidth]{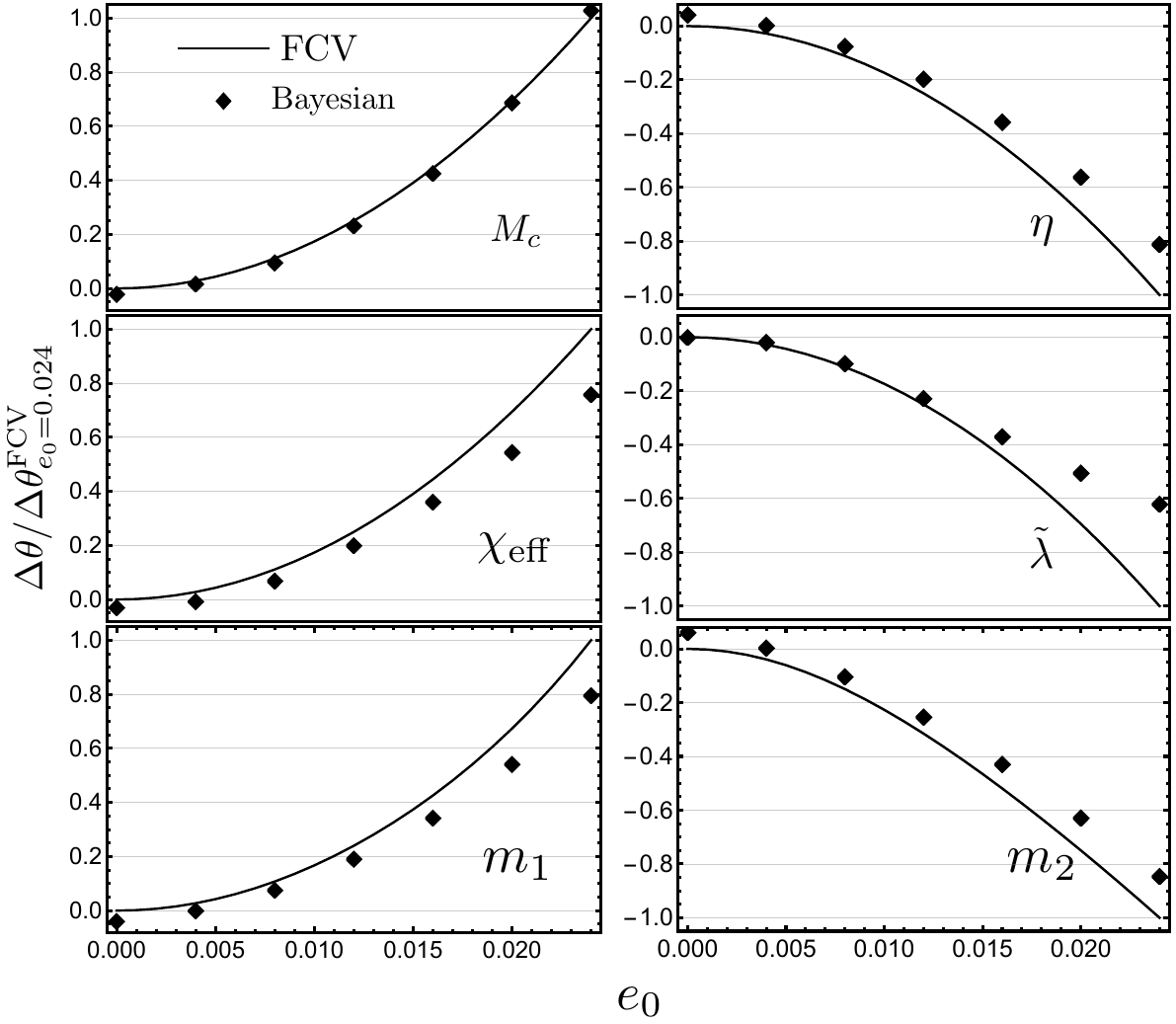}
\caption{\label{fig.bilby-FCV-1d-bias-comparison} Comparison of systematic biases between the FCV and
the Bayesian parameter estimation methods. The bias is normalized by the bias value of FCV at $e_0=0.024$ ($\Delta \theta^{\rm FCV}_{e_0=0.024}$).}
\end{center}
\end{figure}

We compute the FM for the same BNS signal using the same waveform model.
As discussed in the previous work \cite{Cho:2022cdy}, in the FM approach for the BNS system, the measurement errors of all parameters strongly depend on the prior of $\delta \tilde{\lambda}$.
 If we do not incorporate the prior of $\delta \tilde{\lambda}$, the errors are generally much larger than those from the Bayesian parameter estimation results
(for more details, refer to Ref.  \cite{Cho:2022cdy}).
Therefore, we empirically choose the prior as $P_{\delta\tilde{\Lambda}} = 100$  to match the Bayesian PDFs.
A comparison between the Bayesian and FM results is given in Fig. \ref{fig.bilby-Fisher-comparison}.
The gray region corresponds to the Bayesian PDF shown in Fig. \ref{fig.bilby-pdfs}, and the Gaussian function is obtained from the FM.
One can see that the Gaussian functions agree well with the Bayesian PDFs at higher SNRs.
The effectiveness of the FM approach is, in principle, guaranteed in the high SNR limit,
requiring higher SNRs as the number of FM parameters increases.
For a nonspinning case, the FM results are consistent with the Bayesian results at SNRs lower than 50 \cite{Cho:2022cdy}.
 In the subsequent analysis, we assume $\rho=300$  when applying the FM to the FCV method.

Next, we perform Bayesian parameter estimation using noneccentric waveform templates for eccentric signals with $e_0 \leq 0.024$,
and obtain systematic biases for the main parameters $(M_c,\eta,\chi_{\rm eff},\tilde{\lambda})$.
Figure \ref{fig.bilby-1d-bias} shows the Bayesian PDFs for the eccentric signals with $e_0=(0.008, 0.016, 0.024)$.
This result represents how the recovered parameters for eccentric signals can be shifted from their true values (red line).
Even small eccentricities can lead to systematic biases much larger than measurement errors.
The direction of bias is positive for $M_c$ and $\chi_{\rm eff}$, and negative for $\eta$ and $\tilde{\lambda}$.

We also calculate the biases under the same conditions using the FCV method
and compare them with the Bayesian results to examine the reliability of the FCV method.
In Fig. \ref{fig.bilby-FCV-1d-bias-comparison}, we display the biases ($\Delta \theta$) for both the FCV and Bayesian methods in the region of $e_0<0.024$.
Here, to emphasize the similarity between the two methods, the values on the y-axis are given as $\Delta \theta/\Delta \theta^{\rm FCV}_{e_0=0.024}$, 
where $\Delta \theta^{\rm FCV}_{e_0=0.024}$ indicates the bias value of FCV at $e_0=0.024$.
The FCV method appears to be generally reliable as an approximation of the systematic bias for all parameters in the small eccentricity region.
For $M_c$, the FCV curve remarkably well matches the Bayesian results, 
while the FCV overestimates the bias by about $20\%$ at $e_0=0.024$ for $\eta$ and $\chi_{\rm eff}$.
For $\tilde{\lambda}$, the FCV curve is in good agreement with the Bayesian results at low eccentricities ($e_0<0.02$)
but the difference grows up to $40\%$ at $e_0=0.024$ as the Bayesian results deviate from the quadratic pattern.
The results of $(m_1, m_2)$ are obtained from the results of $(M_c,\eta)$,
showing a similar consistency trend as in the case of $\eta$.

\subsection{Monte Carlo study} \label{sec.bias}

\begin{figure}[t]
\begin{center}
\includegraphics[width=\columnwidth]{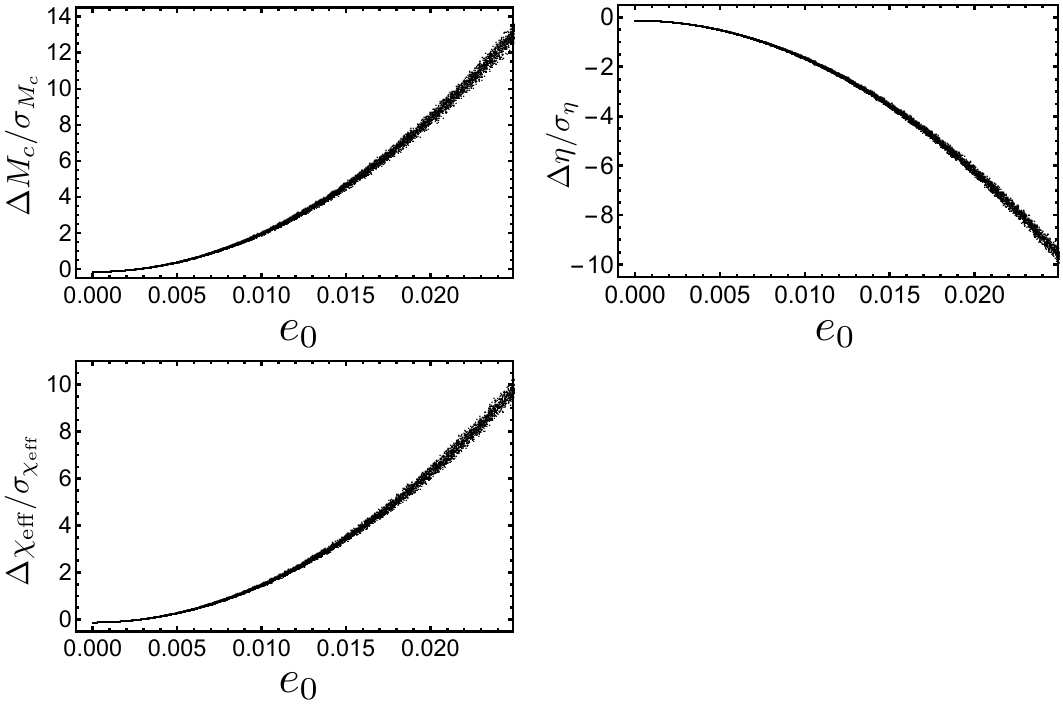}
\caption{\label{fig.monte-carlo-bias}  Distribution of the fractional biases $\Delta \theta/\sigma_{\theta}$ for the $10^4$ Monte Carlo signals.
For the measurement error $\sigma_{\theta}$, we assume $\rho=300$. }
\end{center}
\end{figure}

\begin{figure}[t]
\begin{center}
\includegraphics[width=\columnwidth]{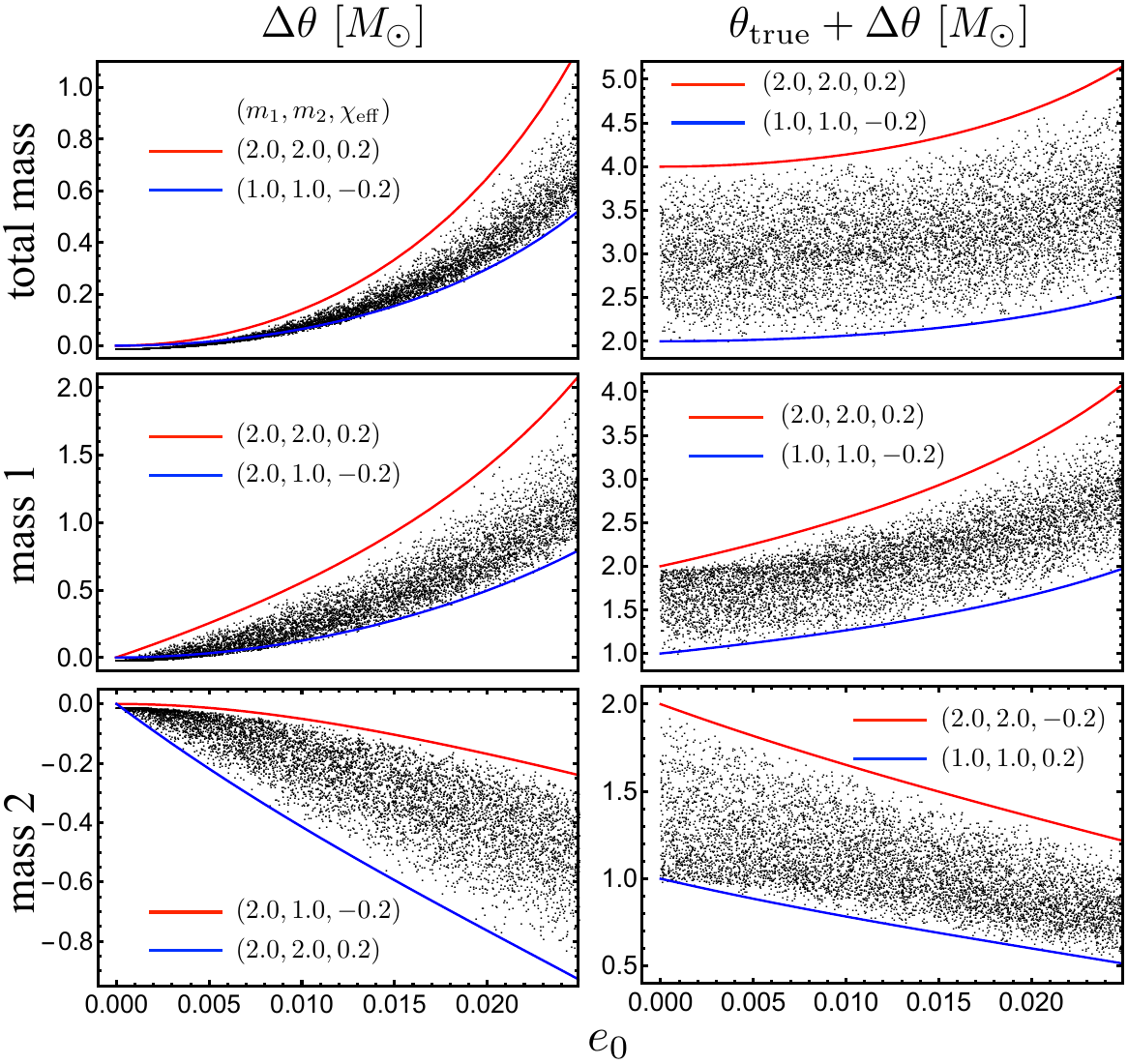}
\caption{\label{fig.monte-carlo-bias-mt-m1-m2}   Distribution of the biases $\Delta \theta$ (left) and the recovered values $\theta_{\rm true}+\Delta \theta$ (right) for the component masses ($m_i$) and the total mass ($M$).
These results are obtained from $\Delta M_c$ and $\Delta \eta$. The red and blue lines represent the upper and lower boundaries of the distribution, respectively.}
\end{center}
\end{figure}

To see the general distribution of bias, we perform a Monte Carlo study using $10^4$ BNS samples.
We randomly generate the signals in the parameter space $m_1$--$m_2$--$\chi_{\rm eff}$--$e_0$ in the ranges of
$1\msun \leq m_{1,2} \leq 2\msun (m_2 \leq m_1)$, $-0.2 \leq \chi_{\rm eff} \leq 0.2$, and $0\leq e_0 \leq 0.024$.
The value of $\tilde{\lambda}$ is determined by applying the APR4 EoS model for all signals.
For these eccentric signals, we calculate systematic biases ($\Delta \theta$) using the FCV method,
and also compute the unbiased measurement errors ($\sigma_{\theta}$) for the same signals that omit the eccentricity.

Figure \ref{fig.monte-carlo-bias} shows the bias distributions for the parameters $(M_c,\eta,\chi_{\rm eff})$.
The impact of bias is better understood when compared to the measurement error,
so we here present the fractional bias ($\Delta \theta/\sigma_{\theta}$).
Note that the measurement error depends on the SNR ($\sigma \propto \rho^{-1}$), but the systematic bias is independent of SNR.
Thus, the value in the y-axis depends on the SNR, and we assume $\rho=300$ in this result.
When $e_0=0.024$,  the magnitude of the fractional bias can increase beyond 10 for $M_c$, and close to 10 for $\eta$ and $\chi_{\rm eff}$.
The vertical width of the distribution band represents the dependence on the mass and spin parameters,
and the thin band indicates that the fractional bias weakly depends on the three parameters in the BNS system.
If the spin range is limited to $-0.05 \leq \chi_{\rm eff} \leq 0.05$ (which corresponds to the low-spin prior range used in the parameter estimation of GW171807 \cite{PhysRevLett.119.161101, PhysRevX.9.011001}), 
the distribution band becomes thinner, which can be roughly expressed as a quadratic fitting curve $\Delta \theta/\sigma_{\theta}=C \times 10^4 e_0^2$.
We find that $C=2.1, -1.55,$ and $1.6$ for $M_c, \eta,$ and $\chi_{\rm eff}$, respectively.
Since the value in the y-axis depends on the SNR ($\Delta \theta/\sigma_{\theta} \propto \rho$),  the fitting coefficient can easily be adjusted according to the SNR,
and the fitting formula can be simply expressed as $\Delta \theta/\sigma_{\theta}=C\rho/300 \times 10^4 e_0^2$.

In addition, we calculate the biases of the component masses and the total mass from the results of $M_c$ and $\eta$.
In Fig. \ref{fig.monte-carlo-bias-mt-m1-m2}, the left panels display the bias ($\Delta \theta$) for the three mass parameters.
The biases are positive and negative for $m_1$ and $m_2$, respectively, and the total mass biases are positive.
The right panels display the recovered parameter values calculated as $\theta_{\rm true}+\Delta \theta$,
and this result shows how far the biased NS mass can deviate from our mass range ($1\msun \leq m_{1,2} \leq 2\msun $) due to the eccentricity.
In this figure, the red and blue lines represent the upper and lower boundaries of the distribution, respectively, and the true values for each line are given in each panel.

\begin{figure}[t]
\includegraphics[width=\columnwidth]{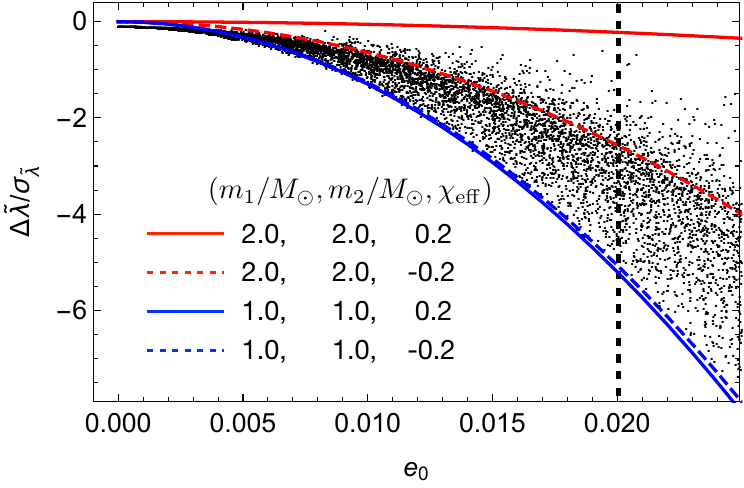}
\caption{\label{fig.monte-carlo-bias-lambda} Monte Carlo results of the fractional bias for the tidal parameter $\tilde{\lambda}$.
We assume $\rho=300$. The red and blue solid lines indicate the upper and lower boundaries, respectively. }
\end{figure}

\begin{figure}[t]
\begin{center}
\includegraphics[width=\columnwidth]{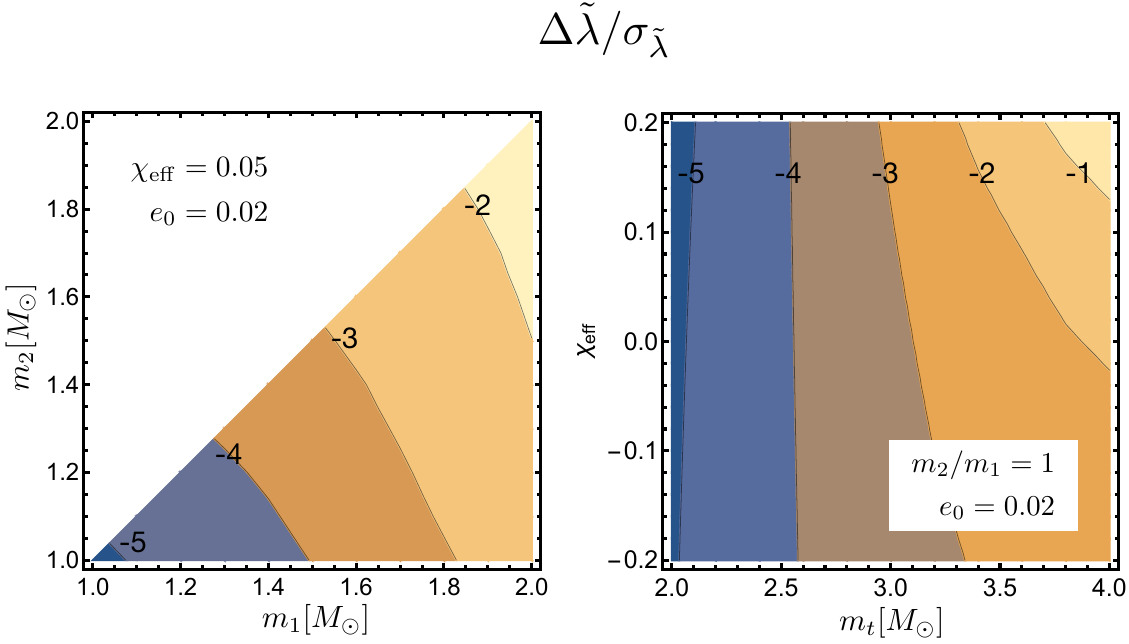}
\caption{\label{fig.bias-2d} Fractional biases $\Delta \tilde{\lambda}/\sigma_{\tilde{\lambda}}$ in the $m_1$--$m_2$ (left panel) and the $m_t$--$\chi_{\rm eff}$ planes (right panel).}
\end{center}
\end{figure}

The Monte Carlo result for the tidal parameter is given separately in Fig. \ref{fig.monte-carlo-bias-lambda}.
Unlike the cases of the other three parameters in Fig. \ref{fig.monte-carlo-bias}, 
the fractional biases ($\Delta \tilde{\lambda}/\sigma_{\tilde{\lambda}}$) are widely distributed according to the mass and spin values.
The upper boundary is obtained from the most massive binary,
the lower boundary is obtained from the lightest binary,
and both the binaries have the largest positive spin value in our spin range ($-0.2 \leq \chi_{\rm eff} \leq 0.2$).
To see specifically the dependence of the bias on the mass and spin,
we display in Fig. \ref{fig.bias-2d} the fractional biases obtained at $e_0=0.02$ in the $m_1$--$m_2$ (left panel) and the $m_t$--$\chi_{\rm eff}$ planes (right panel), respectively.
The spin and the mass ratio are fixed as $\chi_{\rm eff}=0.05$ and $m_2/m_1=1$ in the left and right panels, respectively.
The left plane shows that the magnitude of the fractional bias increases as the total mass decreases
and the dependence on the mass ratio is weak.
The right panel also shows the strong dependence on the total mass over the entire parameter space.
The dependence on the spin is almost negligible in the low mass region, but it becomes stronger with increasing total mass.

\subsection{Application to NS mass inference} 

\begin{figure}[t]
\begin{center}
\includegraphics[width=\columnwidth]{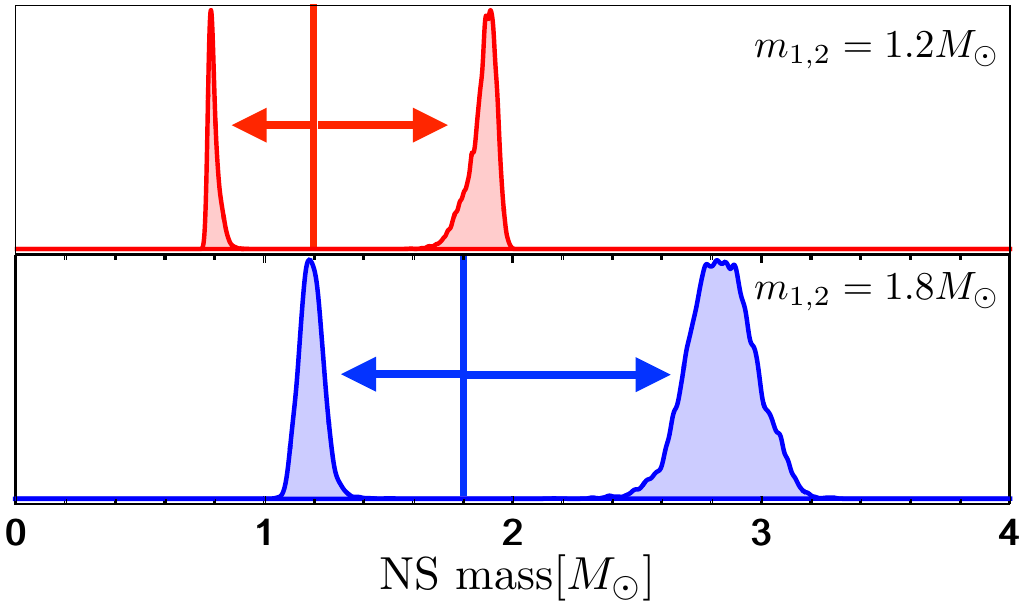}
\caption{\label{fig.mass-bias} Example showing the effect of bias on NS mass inference.
The true values are $m_{1,2}=1.2\msun $ in the upper panel, $1.8 \msun$ in the lower panel, and $(\chi_{\rm eff},e_0, {\rm EoS})=(0.2,0.02, {\rm APR4})$ for both signals.
For all PDFs, the normalization constant has been rescaled to fit in the figure.
This result shows that a BNS signal consisting of two typical NSs can be estimated to be
a BNS signal whose component mass is less than or higher than the typical mass range ($1\msun \leq m_{\rm NS} \leq 2\msun$).}
\end{center}
\end{figure}

It has generally been believed that the NS mass is between $1\msun$ and $2\msun$.
The bias trend of the component NS masses given in Fig. \ref{fig.monte-carlo-bias-mt-m1-m2}
shows that the bias effects due to eccentricity make the recovered masses more asymmetric than the true values.
Here, we show a concrete example of how this trend affects NS mass inference.
We choose two equal-mass signals within the typical NS mass range ($1\msun \leq m_{\rm NS} \leq 2\msun$) with the true values 
$m_{1,2}=1.2\msun $ and $1.8 \msun$, where $(\chi_{\rm eff},e_0, {\rm EoS})=(0.2,0.02, {\rm APR4})$ for both the signals,
and perform Bayesian parameter estimation using noneccentric waveform templates.
The PDFs for the recovered NS masses are given in Fig. \ref{fig.mass-bias} for the signals $m_{1,2}=1.2\msun $ (upper panel)
and $1.8 \msun$ (lower panel), respectively, where the true values are indicated by the vertical lines.
The recovered values are $(m_1,m_2)\simeq (1.9 \msun, 0.8\msun)$ and $(2.8 \msun, 1.2\msun)$, respectively.
This result shows that a BNS signal consisting of two typical NSs can be estimated to be
a BNS signal whose component mass is less than or higher than the typical mass range.
On the other hand, in the FCV Monte Carlo results (right-middle panel in Fig. \ref{fig.monte-carlo-bias-mt-m1-m2}), 
the binary with $(m_1,m_2,\chi_{\rm eff})=(2\msun,2\msun,0.2)$ can have the largest positive bias of the NS mass.
Using this binary source with $e_0=0.02$, we perform Bayesian parameter estimation
and obtain the recovered values as $(m_1,m_2)\simeq (3.2, 1.3)\msun$, even resulting in a NS within the mass gap.

The above results imply that when a BNS signal is observed and its masses are estimated to be asymmetric, 
it is difficult to distinguish whether the true source masses are asymmetric with zero eccentricity 
or symmetric with eccentricity unless eccentric waveform models are used.
In the same way, if a BNS signal is estimated to be exactly symmetric using noneccentric waveforms,
this signal cannot have eccentricity, otherwise the symmetry may have been broken.
Using these bias trends, one may deduce the upper limit of eccentricity from the $e_0$--$\Delta \eta$ relation in Fig. \ref{fig.monte-carlo-bias}.
We here describe a simple example. 
A BNS signal is observed and its properties are estimated as $({\rm SNR}, \eta, \sigma_{\eta})=(300,0.245,0.001)$ using noneccentric waveforms.
If the recovered $\eta$ was biased due to unknown eccentricity buried in the signal,
the maximum bias may have occurred 
when the source was an equal-mass binary (i.e., $\eta_{\rm true}=0.25$), thus $\Delta \eta_{\rm max}=0.245-0.25=-0.005$.
Then, the fractional bias can be $\Delta \eta_{\rm max}/\sigma_{\eta}=-0.005/0.001=-5$,
and this corresponds to $e_0\sim 0.018$ in Fig. \ref{fig.monte-carlo-bias},
consequently, the hidden eccentricity of this signal can be $e_0 < 0.018$.
\footnote{For signals with arbitrary SNRs and errors ($\sigma_{\eta}$), the same procedure can be applied after adjusting the errors
to those when the SNR is 300. Note that the bias ($\Delta \eta$) is independent of the SNR.}
Therefore, the true values of the signal may have been between the combinations $(\eta, \sigma_{\eta})=(0.25,0.018 )$ and $(0.245,0.0)$.

\subsection{Application to NS EoS inference}

\begin{figure}[t]
\begin{center}
\includegraphics[width=\columnwidth]{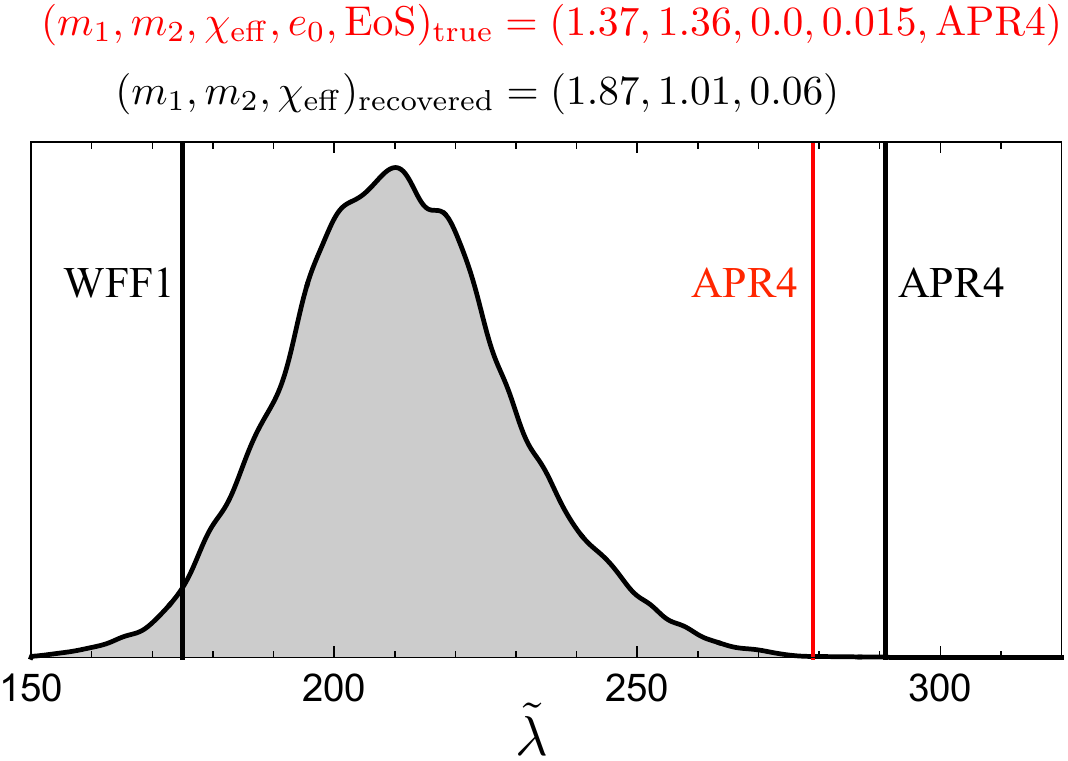}
\caption{\label{fig.eos-bias-1d} Example showing that the recovered Bayesian result prefers the WFF1 EoS model over the true model APR4.
The true and recovered values are shown at the top.  The red vertical line indicates the true value of $\tilde{\lambda}$ for the true signal $(1.37\msun,1.36\msun)$,
and the black vertical lines correspond to the values of $\tilde{\lambda}$
obtained by applying the recovered masses $(1.87\msun,1.01\msun)$ to the EoS models WFF1 and APR4, respectively. }   
\end{center}
\end{figure}

The bias on the tidal parameter directly affects the NS EoS prediction.
Since the EoS is determined by the value of $\tilde{\lambda}$, incorrect estimation of $\tilde{\lambda}$ due to bias
results in incorrect EoS choices among various theoretical models.
Here, we describe a specific case in which the recovered tidal value prefers another EoS model rather than the (assumed) true model.
We select an eccentric BNS source with the true values 
as $(m_1,m_2,\chi_{\rm eff},e_0)=(1.37\msun,1.36\msun,0.0,0.015)$, which are similar to those of  GW170817 except for the eccentricity.
We assume that the true value of the tidal parameter is determined according to the APR4 EoS model,
 and thus $\tilde{\lambda}_{\rm true}=279$ for this signal.
We perform Bayesian parameter estimation using noneccentric waveform templates
and obtain the recovered values as $(m_1,m_2,\chi_{\rm eff})=(1.87\msun,1.01\msun,0.06)$.
We show the recovered PDF of $\tilde{\lambda}$ separately in Fig. \ref{fig.eos-bias-1d}.
The red vertical line indicates the true value of $\tilde{\lambda}$ for the true signal,
and the black vertical lines correspond to the values of $\tilde{\lambda}$
obtained by applying the recovered masses $(1.87\msun,1.01\msun)$ to the EoS models WFF1 and APR4, respectively.
This example exactly illustrates that the recovered Bayesian result prefers the WFF1 EoS model over the true model APR4.

\begin{figure}[t]
    \centering
        \includegraphics[width=\linewidth]{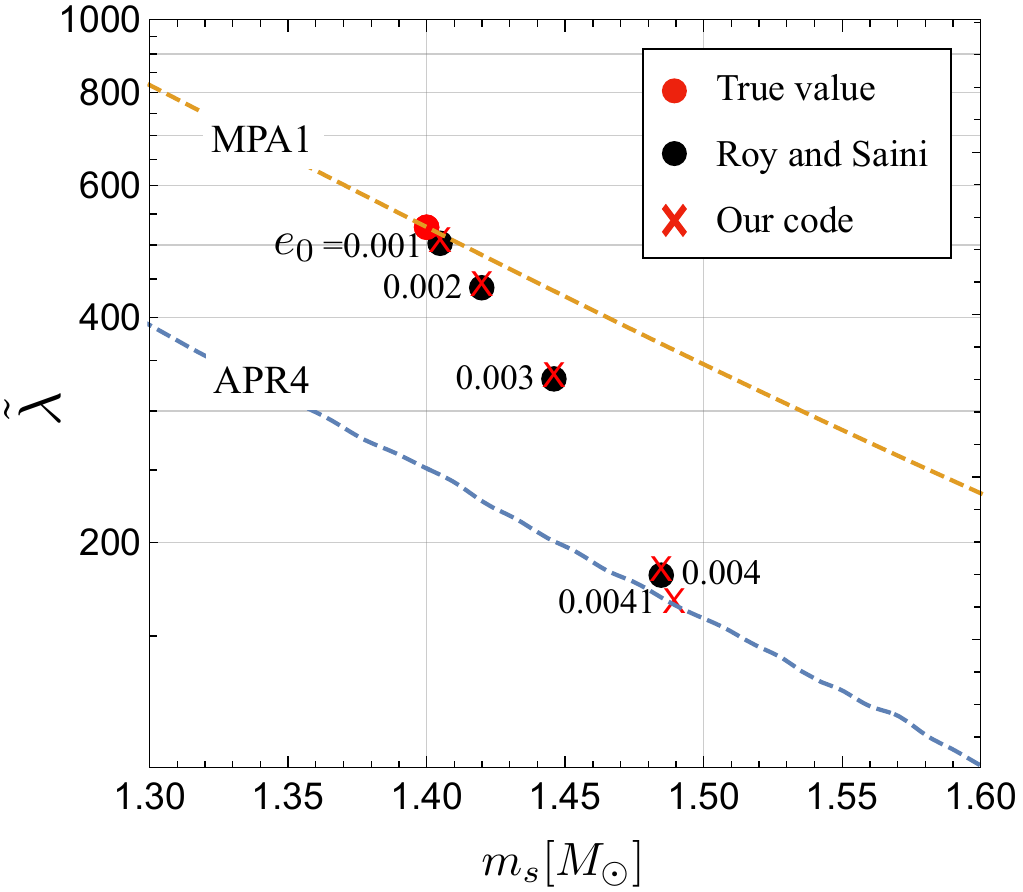}
        \caption{Bias due to eccentricity in the $m_s$--$\tilde{\lambda}$ plane, where $m_s$ indicates the source frame NS mass. 
        The true position of the signal and its recovered positions are marked in red and black dots, respectively.
         The upper and lower dashed lines follow the MPA1 and APR4 EoS models, respectively.
         This example illustrates that parameter estimation using noneccentric waveforms for the ($1.4\msun,1.4\msun$) eccentric BNS signal whose EoS follows the MPA1 model
can derive the result preferring the APR4 model when $e_0\sim0.0041$.}
         \label{fig:eos-comparison}
         \end{figure}

Roy and Saini \cite{DuttaRoy:2024aew} also studied the systematic errors due to unmodeled eccentricity in the tidal deformability measurement and its implications for NS EoS,
taking into account the 3 G detectors, Cosmic Explorer and Einstein Telescope.
They employed the same FM and FCV methods, but considered a different parameter set as $(M_c,\eta,\chi_1,\chi_2,\tilde{\lambda},\phi_c,t_c)$ for the FM.
They used the component spin parameters instead of the single spin parameter $\chi_{\rm eff}$,
and omitted the second tidal parameter $\delta \tilde{\lambda}$ because it vanishes in equal-mass systems. 
Consequently, the number of parameters is seven, the same as in our FM.
For cross-checking, we modify our FM and FCV codes, consisting of the same parameter set as in Ref. \cite{DuttaRoy:2024aew}
and calculate the bias for the same BNS source $(m_{1,2},\chi_{1,2},{\rm EoS})=(1.4\msun,0.05,{\rm MPA1})$ under the same conditions 
(i.e., PSD, $f_{\rm min}, f_{\rm max}$, etc. For more details, refer to Ref. \cite{DuttaRoy:2024aew}.).
In Fig. \ref{fig:eos-comparison}, the true position of the signal and its recovered positions are marked in red and 
black dots in the $m_s$--$\tilde{\lambda}$ plane 
\footnote{In this result, Roy and Saini \cite{DuttaRoy:2024aew} ignored the bias effects on $\eta$ and assumed that the systematic error
is the same on both the NS in the binary, i.e., $m_s=\Delta M/2$.} (where $m_s$ indicates the source frame NS mass),  respectively.
We find that the result of the biased positions agrees well with that of \cite{DuttaRoy:2024aew}
(cf. Fig. 3 of Ref. \cite{DuttaRoy:2024aew}).
We recalculate the biases (marked in a red cross) using our original FM and FCV codes consisting of the parameter set $(M_c,\eta,\chi_{\rm eff},\tilde{\lambda}, \delta\tilde{\lambda},\phi_c,t_c)$
and find no difference from the results of \cite{DuttaRoy:2024aew}.
On the other hand, the upper and lower dashed lines follow the MPA1 and APR4 EoS models, respectively.
Therefore, this example illustrates that parameter estimation using noneccentric waveforms for the ($1.4\msun,1.4\msun$) eccentric BNS signal whose EoS follows the MPA1 model
can derive the result preferring the APR4 model when $e_0\sim0.0041$.

\section{Summary and discussion}

We investigated the impact of eccentricity on GW parameter estimation with the aLIGO detector sensitivity for spinning BNS signals
and demonstrated that even small eccentricity ($e_0\leq 0.024$ at 10 Hz) buried in the signal 
can shift the mass, spin, and tidal parameters significantly from their true values.
As in our previous work \cite{Cho:2022cdy}, we employed the FM and FCV methods 
to calculate the measurement error and the systematic bias, 
and those were validated by comparing with the Bayesian parameter estimation results.
The main difference between our two works is the opposite direction of bias for the mass parameters except for $M_c$.
In particular, for $\tilde{\lambda}$, the direction is reversed from positive to negative due to the inclusion of the spin parameter,
and this implies that the recovered tidal parameter favors a stiffer EoS than the true EoS for the nonspinning case, but a softer EoS for the spinning case.
The bias directions are summarized in Table. \ref{tab1}.

\begin{table}[t]
 \begin{ruledtabular}
\begin{tabular}{cccccccc}
Parameter&$M_c$ & $\eta$ & $m_1$  & $m_2$ & $M$ &$\chi_{\rm eff}$&$\tilde{\lambda}$  \\
\hline
Nonspinning & $+$  &$ +$  &$-$ & $ +$ & $-$ &...&$+$\\
Spinning & $+$ & $-$ & $+$ & $-$ & $+$  &$+$&$-$\\
\end{tabular}
    \end{ruledtabular}
\caption{Direction of bias due to eccentricity. For $\tilde{\lambda}$, the positive ($+$) and negative ($-$) directions imply
that the recovered tidal parameter favors a stiffer EoS and a softer EoS than the true EoS, respectively.}
\label{tab1}
\end{table}

Using the $10^4$ Monte Carlo BNS signals distributed in the parameter space $m_1$--$m_2$--$\chi_{\rm eff}$--$e_0$,
we calculated the systematic biases ($\Delta \theta$)
for the chirp mass ($M_c$), symmetric mass ratio ($\eta$), effective spin ($\chi_{\rm eff}$), and effective tidal deformability ($\tilde{\lambda}$),
and showed their generalized distributions using the quantity, fractional bias $\Delta \theta / \sigma_{\theta}$.
We also showed the distributions of bias and recovered value for the component masses and total mass.
We found that the fractional bias weakly depends on the three main parameters  $M_c, \eta,$ and $\chi_{\rm eff}$,
displaying a narrow band increasing or decreasing quadratically with increasing $e_0$,
while the biases of $\tilde{\lambda}$ showed a wide distribution depending on the values of the mass and spin parameters at a given $e_0$.

We applied our result of bias to the inference of neutron star properties 
and performed Bayesian parameter estimation for specific cases to identify important implications.
The bias effects due to eccentricity always make the recovered component masses more asymmetric than the true masses.
We showed that a BNS signal consisting of two neutron stars within the typical mass range $[1,2]\msun$ can be estimated to be
a BNS signal whose component mass is much smaller or much larger than the typical mass range, 
even up to the mass gap for a specific case $(m_1,m_2,\chi_{\rm eff},e_0)_{\rm true}=(2\msun,2\msun,0.2,0.02)$.
In particular, we found that parameter estimation using noneccentric waveforms for eccentric
BNS signals can yield false predictions for the neutron star equation of state.
We presented two concrete Bayesian examples: i) the recovered Bayesian PDF for $\tilde{\lambda}$ prefers the WFF1 EoS model
over the true model APR4, ii) parameter estimation using noneccentric waveforms for the ($1.4\msun,1.4\msun$) eccentric BNS signal whose EoS follows the MPA1 model
can derive the result preferring the APR4 model when $e_0\sim0.0041$.

The TaylorF2 model is a purely analytic PN model, which has a very simple dependence of the amplitude on frequency.
This model only describes the inspiral phase and is inaccurate when the binary is close to merger.
For merging BNSs, the merger happens at frequencies above 1 kHz due to their low masses, 
and the aLIGO is less sensitive in this frequency region.
Therefore, the TaylorF2 model has generally been considered suitable for BNS signals.
The TaylorF2 model, along with three high-precision models, was used for parameter estimation of GW170817,
and the result showed marginal differences from the results of the three models \cite{PhysRevX.9.011001}.
The effectiveness of TaylorF2 may have to be comprehensively investigated for future detectors
 whose sensitivity will be improved in the high frequency region.

The tidal parameter $\delta \tilde{\lambda}$ first appears at 6PN order, and this parameter has a negligible contribution to the GW phase compared to $\tilde{\lambda}$. 
We found that this parameter cannot be well recovered even at $\rho \simeq 1000$ using BILBY.
Huez  \etal \cite{huez2025k}  have also shown that
even with the next generation detector, Einstein Telescope, $\rho > 1000$ will be required to confidently
constrain $\delta \tilde{\lambda}$.
In the FM, $\delta \tilde{\lambda}$ is almost unmeasurable ($\sigma_{\delta \tilde{\lambda}} > 10^4$) 
unless appropriate prior information is incorporated \cite{Cho:2022cdy}.
Thus, in this work, we chose the prior of $\delta \tilde{\lambda}$ empirically with no theoretical justification for that choice.
On the other hand, we included $\delta \tilde{\lambda}$ as a free parameter in our analysis for consistency with the Bayesian parameter estimation.
However, we confirmed that our results were nearly unchanged even though we omitted this parameter from the FM component.
A comprehensive study of how to effectively deal with this parameter in the FM approach would be necessary.

The FM method has been widely used in past parameter estimation studies due to its ease of construction and application.
However, as described in this work, it is important to verify the reliability of the method before its practical application,
which can be done simply by comparing it to the Bayesian parameter estimation method \cite{Cho_2022,PhysRevD.106.084056}.
The FCV method can also be verified by comparing Bayesian parameter estimation methods.
 On the other hand, the FM and FCV methods will be much more efficient for the 3G detectors
 because longer signal lengths require much longer computational time for parameter estimation.
The verification of the accuracy of the two methods is also important in this regard,
and our methodology used in this work is easily applicable to 3G detector sensitivity curves,
which will be our future work.

\newpage
\newpage
\newpage

\section*{Acknowledgments}
This work was supported by the National Research Foundation of Korea (NRF) grants funded by the Korea government 
(No. RS-2023-NR076639, No. RS-2023-00242247, and No. RS-2023-00301938).

\bibliographystyle{apsrev4-2}

\bibliography{biblio}

%\end{references}

\end{document}